\documentclass[twocolumn]{webofc}
\usepackage[varg]{txfonts}   
\woctitle{LHCP 2013}

\newcommand{\muSUSY}{\mu_{\rm susy}}
\newcommand{\softsusy}{SOFTSUSY}
\newcommand{\mgluino}{m_{\tilde{g}}}
\newcommand{\mstopone}{m_{\tilde{t}_1}}
\newcommand{\mstoptwo}{m_{\tilde{t}_2}}
\newcommand{\msquark}{m_{\tilde{q}}}

\begin{document}
\title{Higgs boson production in the SM and MSSM to NNLO and beyond
\\[-3em]\mbox{}\hfill
{\small SFB/CPP-13-40, TTP13-022}
\\[0.7em]
}
%
%

\author{Matthias Steinhauser\inst{1}\fnsep\thanks{\email{matthias.steinhauser@kit.edu}}
}

\institute{Karlsruhe Institute of Technology (KIT), Institut f\"ur Theoretische Teilchenphysik
          }

\abstract{%
  In this contribution a brief review about the status of 
  higher order corrections to Higgs boson production 
  within the Minimal Supersymmetric Standard Model (MSSM)
  is given. Furthermore the first activities towards 
  third-order corrections in the Standard Model (SM) are discussed.
}
\maketitle
%


\section{Introduction}
\label{sec::intro}
The discovery of a new Higgs boson-like particle at LHC~\cite{:2012gk,:2012gu}
has triggered plenty activities with the aim to pin down its properties
like couplings and decay rates.
In this contribution we consider the production cross section
of a Higgs boson in the gluon-fusion channel
and discuss the status both for the SM and the MSSM.

The theoretical framework of our calculations is the effective theory where
all particles which are heavier than the Higgs boson are integrated out.
This leads to an effective Higgs-gluon interaction which is described by
\begin{eqnarray}
  {\cal L}_{\rm eff} &=& -\frac{H}{v} C_1 \frac{1}{4} G_{\mu\nu} G^{\mu\nu}
  \,,
  \label{eq::leff}
\end{eqnarray}
where $G_{\mu\nu}$ is the gluon field strength tensor and $C_1$ is the
coupling (or matching coefficient) containing the remnant dependence on the
heavy degrees of freedom.  Within the SM $C_1$ only depends on the top quark
mass via $\ln(\mu^2/m_t^2)$ where $\mu$ is the renormalization scale.
In the MSSM, $C_1$ becomes a complicated function of all heavy mass scales and
$\mu$.

In Refs.~\cite{Harlander:2009mq,Pak:2009dg,Harlander:2009my,Pak:2011hs} it has
been demonstrated that at next-to-next-to-leading order (NNLO)
the effective-theory approach of Eq.~(\ref{eq::leff}) approximates the exact
SM result with an accuracy below 1\%, in particular for Higgs boson masses
around $126$~GeV.
Numerical NLO calculations~\cite{Anastasiou:2008rm} 
suggest a similar behaviour in the MSSM.

In Section~\ref{sec::mssm} NNLO SUSY QCD corrections are considered within the
MSSM. Afterwards, we summarize in Section~\ref{sec::sm} the first steps
towards N$^3$LO in the SM.


\section{\label{sec::mssm}NNLO corrections to $gg\to H+X$ in the MSSM}

The computation of higher order supersymmetric corrections within the
effective-theory framework requires the evaluation of loop corrections to the
matching coefficient $C_1$. Several groups have computed two-loop corrections
both in the top~\cite{Harlander:2003bb,Harlander:2004tp,Degrassi:2008zj} and
bottom sector~\cite{Anastasiou:2008rm,Degrassi:2010eu,Harlander:2010wr}.  NLO
calculations in the full theory have been performed in
Ref.~\cite{Anastasiou:2008rm}.  Building blocks for a (semi) analytic
full-theory calculation have been provided in
Refs.~\cite{Muhlleitner:2006wx,Bonciani:2007ex}; a complete calculation along
these lines is still missing.  Recently the effective-theory NLO corrections
have been implemented in the publicly available computer code {\tt
  SusHi}~\cite{Harlander:2012pb}. At NNLO the rough approximation of
Ref.~\cite{Harlander:2003kf} has been implemented, i.e., the genuine SUSY
corrections to $C_1$ have been set to zero at three loops.

\begin{figure}
  \centering
  \includegraphics[width=1.\linewidth]{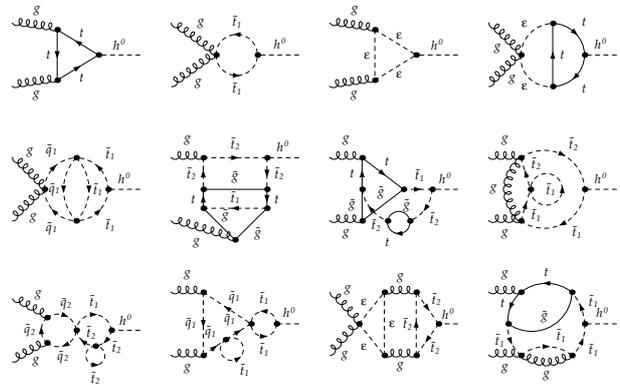}
  \caption{Feynman diagrams contributing to $C_1$.}
  \label{fig::diags1} 
\end{figure}

In this contribution we want to discuss the numerical effect of three-loop
corrections to $C_1$ which are needed in order to obtain a complete NNLO
prediction of the Higgs boson production cross section.  Sample diagrams
contributing to $C_1$ at one, two and three loops are shown in
Fig.~\ref{fig::diags1}.  The symbols $t$, $\tilde{t}_i$, $g$, $\tilde{g}$, $h$
and $\varepsilon$ denote top quarks, top squarks, gluons, gluinos, Higgs
bosons and $\varepsilon$ scalars, respectively.\footnote{$\varepsilon$ scalars
  are auxiliary particles introduced to implement regularization by
  dimensional reduction~\cite{Siegel:1979wq} which respects supersymmetry.}
For the computation of 
$C_1$ it is possible to expand the Feynman integral in the external gluon
momenta which leads to vacuum diagrams. In contrast to the SM, in the MSSM
many different mass scales are present which increases the complexity of the
calculation significantly. In fact, the currently available tools do not allow
for an exact calculation and one has to rely on approximation methods. In
Refs.~\cite{Pak:2012xr,Kurz:2012ff} hierarchies in the occurring masses have
been defined in such a way that phenomenological interesting scenarios can be
studied. Furthermore, sophisticated expansion schemes have been defined which
select the representation of the perturbative series with 
largest radius of convergence.
Moreover, reliable error estimates for each point in the parameter
space are obtained.  In this way three-loop corrections to $C_1$ have been
evaluated for the top and bottom sector, neglecting, however, the bottom
Yukawa coupling.  In a first step a simplified scenario where all
supersymmetric masses are identical has been considered
in~\cite{Pak:2010cu}. In Ref.~\cite{Pak:2012xr} the results have been
generalized by considering various hierarchies of the involved supersymmetric
particle masses.  Furthermore, details on the renormalization procedure and
the treatment of evanescent couplings are discussed. The results
of~\cite{Pak:2012xr} have been cross-checked in Ref.~\cite{Kurz:2012ff} where
a low-energy theorem has been used in order to obtain $C_1$ from the decoupling
constant of $\alpha_s$.

In order to incorporate all know results in the 
final prediction for the cross section we apply the 
following formula
\begin{eqnarray}
  \label{eq::sigma}
  &&\sigma(pp\to h+X) = \left( 1 + \delta^{\rm EW} \right) 
  \times
  \nonumber  \\&&
  \Bigg[\sigma_{tb}^{\rm SQCD}(\mu_s)\bigg|_{\rm NLO}
    -   \sigma_t^{\rm SQCD}(\mu_s)\bigg|_{\rm NLO}
  \nonumber  \\&&
    +   \sigma_t^{\rm SQCD}(\mu_s,\mu_h)\bigg|_{\rm NNLO}
  \Bigg]
  \,,
\end{eqnarray}
where $\sigma_{tb}$ refers to the NLO result including all top and bottom
effects. After subtracting the top quark/top squark contributions with the 
help of $\sigma_t^{\rm SQCD}(\mu_s)|_{\rm NLO}$ we can add the result
from the 
top quark/top squark up to NNLO.\footnote{Note that $\sigma_t$ also contains
  contributions from a non-vanishing Higgs-bottom squark coupling, see
  Ref.~\cite{Pak:2012xr} for details. Since they are small we
  refer in Eq.~(\ref{eq::sigma}) only to the top quark/top squark sector.} 
Finally, electroweak effects are taken into account
in a multiplicative way.

\begin{figure}
  \centering
  \includegraphics[width=1.0\linewidth]{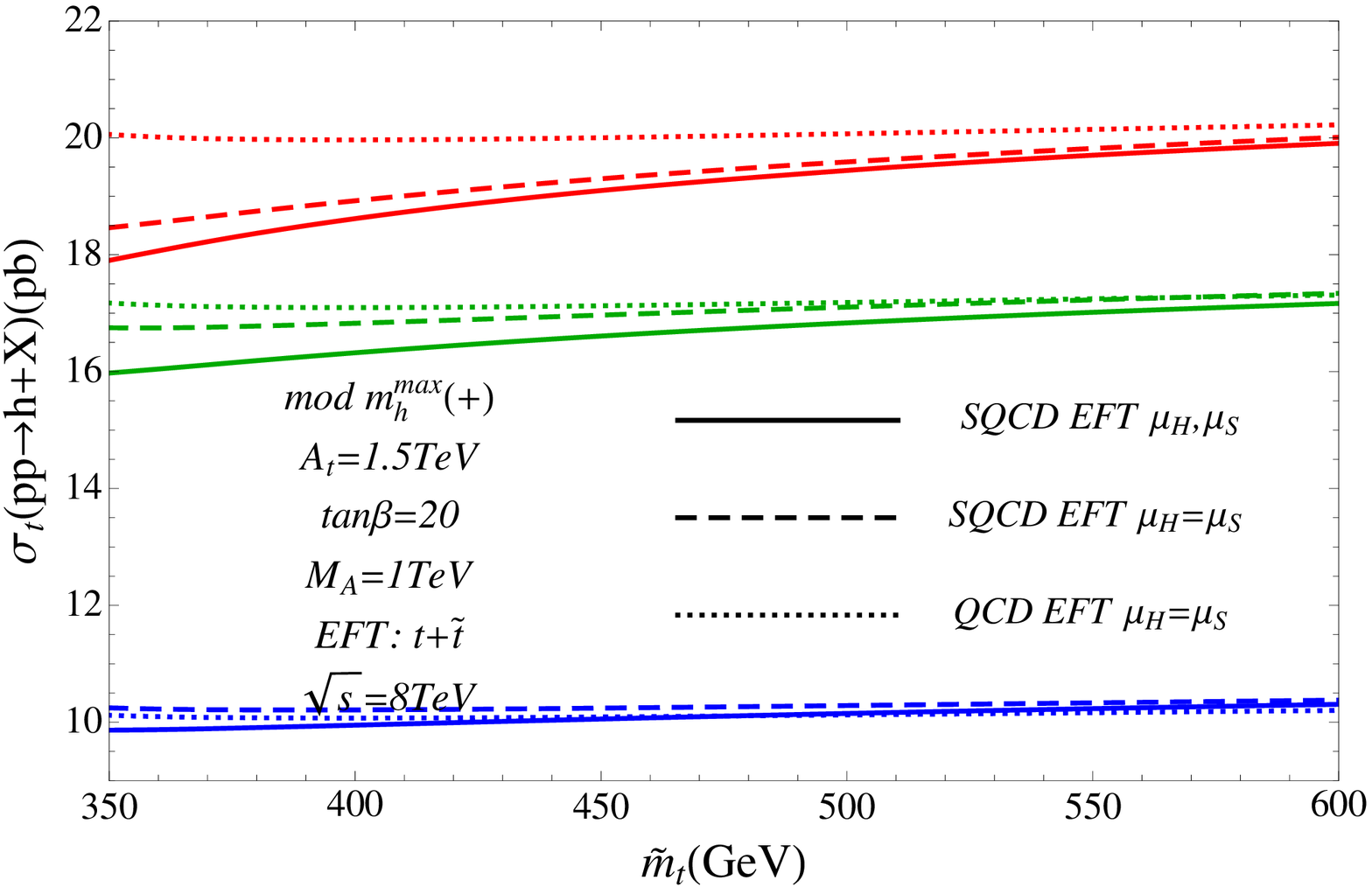}
  \\ (a) \\[2em]
  \includegraphics[width=1.0\linewidth]{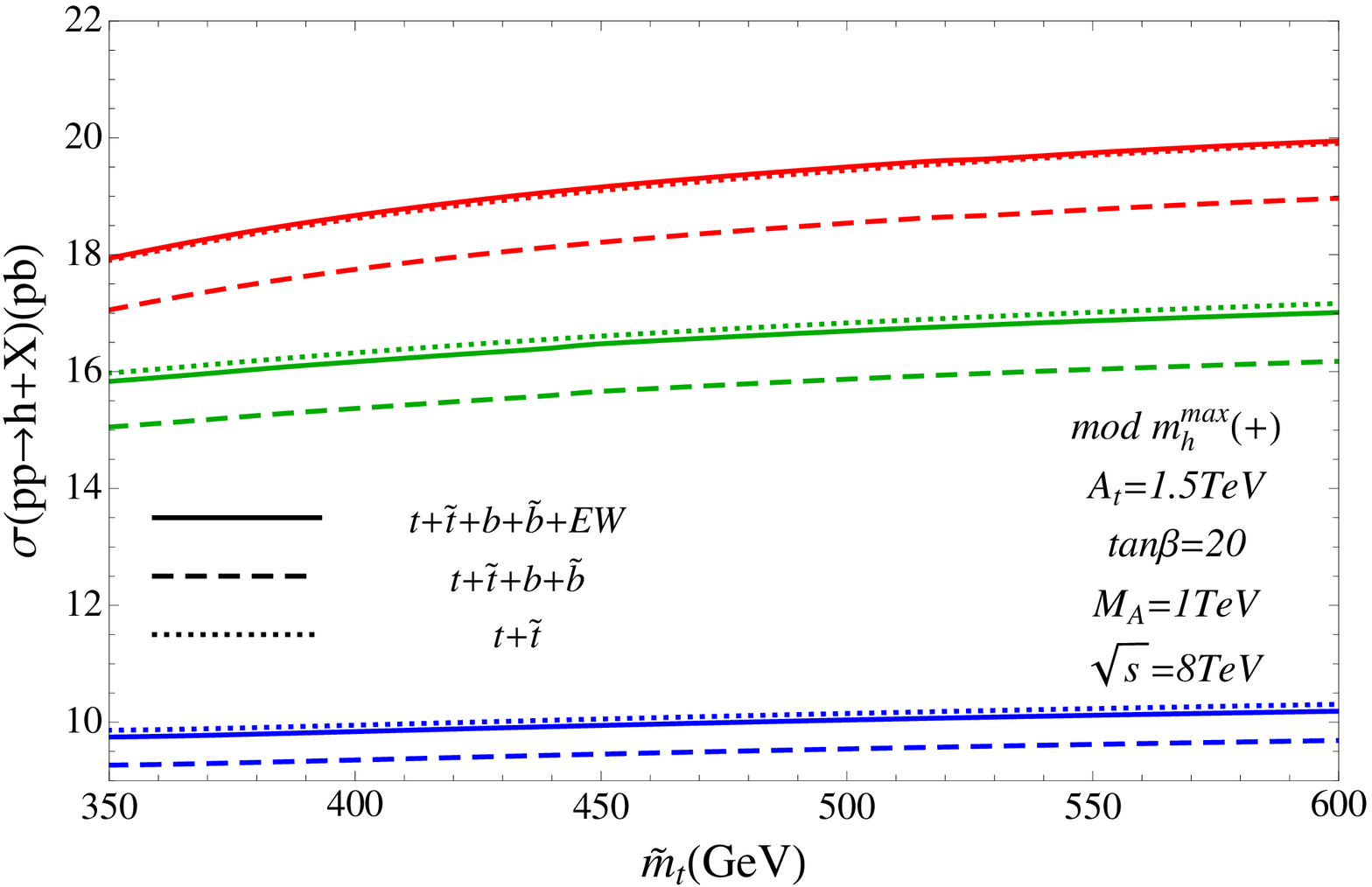}
  \\ (b)
  \caption{Cross section as a function of the 
    singlet soft SUSY breaking parameter of the                                        
    right-handed top squark, $\tilde{m}_t$. (a) top quark/top squark
    contribution $\sigma_t$. (b) complete contribution including
    also bottom quark and electroweak effects as 
    described in Eq.~(\ref{eq::sigma}).}
  \label{fig::sig_mssm}
\end{figure}

In Fig.~\ref{fig::sig_mssm} we discuss numerical effects of the individual
terms in Eq.~(\ref{eq::sigma}) using the $m_h^{\rm max}$ scenario of
Ref.~\cite{Carena:2002qg} as a basis. We apply slight modifications which
lead to the following parameters (see Ref.~\cite{Pak:2012xr} for explanations
of the parameters)
\begin{align}
  &&A_b=A_\tau=2469.48~\mbox{GeV}\,,
  &&A_t=1500~\mbox{GeV}\,,
  \nonumber\\
  &&M_1 = 5 s_W^2 / (3 c_W^2) M_2\,,
  &&M_2 = 200~\mbox{GeV}\,,
  \nonumber\\
  &&M_3 = 800~\mbox{GeV}\,,
  &&M_A=1000~\mbox{GeV}\,,
  \nonumber\\
  &&\muSUSY=200~\mbox{GeV}\,,
  &&m_{\rm susy}=1000~\mbox{GeV}\,,
  \nonumber\\
  &&\tan\beta = 20\,.
  \label{eq::parameters3}
\end{align}
In addition we have the parameter
$\tilde{m}_t$, the singlet soft SUSY breaking parameter of the
right-handed top squark, which is varied in Fig.~\ref{fig::sig_mssm}.
The default value $\tilde{m}_t=400$~GeV in combination with
{\tt \softsusy}~\cite{Allanach:2001kg} leads to the following values for the
$\overline{\rm DR}$ masses
\begin{align}
  & \mstopone = 370~\mbox{GeV}\,, & \mstoptwo = 1045~\mbox{GeV}\,, & 
  \nonumber\\
  & \msquark = 1042~\mbox{GeV}\,, & \mgluino = 860~\mbox{GeV}\,, & 
\end{align}
where $\msquark$ corresponds to the average of
$m_{\tilde{u}}$, $m_{\tilde{d}}$, $m_{\tilde{s}}$, $m_{\tilde{c}}$ and $m_{\tilde{b}}$
and the renormalization scale has been set to the on-shell top quark mass.

In Ref.~\cite{Pak:2012xr} the program {\tt H3m}~\cite{Harlander:2008ju,Kant:2010tf}
has been used in order to compute the lightest MSSM Higgs boson mass.
Combining {\tt H3m} with version~2.6.5 of {\tt FeynHiggs}~\cite{Frank:2006yh} 
and version~3.1.1 of {\tt \softsusy}~\cite{Allanach:2001kg}
leads to a Higgs boson mass of approximately 126~GeV
almost independent of $\tilde{m}_t$~\cite{Pak:2012xr}.

In Fig.~\ref{fig::sig_mssm}(a) the quantity $\sigma_t^{\rm SQCD}$ is shown as
a function of $\tilde{m}_t$ at LO, NLO and NNLO (from bottom to top). For each
order three curves are shown where the dotted curve corresponds to the SM. The
SQCD corrections are included in the dashed and solid line where for the
former the soft and hard renormalization scales, $\mu_s$ and $\mu_h$ have been
identified with $M_h/2$ and for the latter $\mu_s=M_h/2$ and $\mu_h=M_t$ has
been chosen.

One observes that the difference between SM and MSSM becomes small for
increasing $\tilde{m}_t$ which is expected since in this limit the spectrum
becomes heavy.  However, for smaller values of $\tilde{m}_t$ a
sizeable effect of the generic SUSY contribution is visible. For example, for
$\tilde{m}_t=400$~GeV a reduction of the SM cross section of about 5\% is
observed when including NNLO supersymmetric corrections.

The difference between the dashed and solid line in Fig.~\ref{fig::sig_mssm}(a)
quantifies the effect of the resummation of $\ln (M_h^2/M^2_{\rm heavy})$
where $M_{\rm heavy}$ is a heavy mass scale present in the calculation of
$C_1$. It is negligible for
large $\tilde{m}_t$, however, for smaller values it can lead to a visible
effect.

It is interesting to note that supersymmetric three-loop
corrections to $C_1$ computed in Ref.~\cite{Pak:2012xr,Kurz:2012ff} provide an important
contribution to the difference of the solid and dotted curve in
Fig.~\ref{fig::sig_mssm}(a). In fact, if we choose $\tilde{m}_t=400$~GeV 
and identify the three-loop coefficient with the SM one
a reduction of only 3\% and not 5\% is observed.

Let us finally present results for $\sigma(pp\to h+X)$ which include in
addition bottom quark contributions up to NLO and furthermore also electroweak
corrections.  In Fig.~\ref{fig::sig_mssm}(b) we show the dependence on
$\tilde{m}_t$ at LO, NLO and NNLO (from bottom to top).  The dotted curves in
Fig.~\ref{fig::sig_mssm}(b) correspond to the solid ones of
Fig.~\ref{fig::sig_mssm}(a), i.e. they only include the top-sector
contribution. The inclusion of the bottom quark effects at NLO
(cf. Eq.~(\ref{eq::sigma})) leads to a reduction of about 5\% as shown by the
dashed curves. The reduction is basically independent of $\tilde{m}_t$ and
$\tan\beta$.\footnote{The dependence on $\tan\beta$ is studied in
  Ref.~\cite{Pak:2012xr}.}  Thus, even for $\tan\beta=20$ the bottom quark effects
are small for the considered scenarios and, hence, at NNLO the approximation
$m_b=0$ is justified.  The reduction due to bottom quark effects is to a large
extend compensated by the electroweak corrections taken into account
multiplicatively as can be seen by the solid line
which includes all contributions of Eq.~(\ref{eq::sigma}).

To conclude this section let us remark that the Higgs boson production cross
section within the MSSM is known to the same accuracy as in the SM.  The
supersymmetric NNLO corrections can effect the production cross section by a
few percent in case there is a splitting in the top squark masses by a few
hundred GeV and the overall scale of the spectrum is not too heavy. Such
effects are certainly relevant once the experimental precision for the cross
section measurement is considerably below 10\%, in particular, once there are
hints for new particles from direct searches for supersymmetry.

\section{\label{sec::sm}First steps towards N$^3$LO in the SM}

The cross section for Higgs boson production in gluon fusion has been computed
to NLO~\cite{Dawson:1990zj,Spira:1995rr} and 
NNLO~\cite{Harlander:2002wh,Anastasiou:2002yz,Ravindran:2003um,Marzani:2008az,Harlander:2009mq,Pak:2009dg}.\footnote{We refer to the reports of the LHC Higgs cross section
  working groups for further details and extended lists of references~\cite{Dittmaier:2011ti,Dittmaier:2012vm}.} Nevertheless
the contribution from unknown higher orders is estimated to be of the order of
10\% which asks for a N$^3$LO
calculation.
Different groups have started to look into this issue which shall be briefly
summarized in the following:
\begin{itemize}
\item In Ref.~\cite{Chetyrkin:1997un} the four-loop corrections to the
  matching coefficient $C_1$ have been constructed from the three-loop
  decoupling constant for the strong coupling constant with the help of
  renormalization group methods and a low-energy theorem. In
  Refs.~\cite{Schroder:2005hy,Chetyrkin:2005ia} the result has been confirmed
  by an explicit calculation of the four-loop decoupling constant.
\item The three-loop corrections to the 
  massless Higgs-gluon form factor have been obtained by two 
  independent calculations~\cite{Baikov:2009bg,Gehrmann:2010ue}
  (see also Ref.~\cite{Lee:2010cga}).
\item The ${\cal O}(\epsilon)$ contributions to the master integrals needed
  for the NNLO calculation has been computed in Refs.~\cite{Pak:2011hs,Anastasiou:2012kq}.
\item Results for the LO, NLO and NNLO partonic cross sections expanded up to
  order $\epsilon^3$, $\epsilon^2$ and $\epsilon^1$, respectively, 
  have been obtained in Ref.~\cite{Hoschele:2012xc}.
\item All contributions from convolutions of partonic cross sections 
  with splitting functions, which are needed for the complete
  N$^3$LO calculation, are provided in Ref.~\cite{Hoschele:2012xc}.
  The results of~\cite{Hoschele:2012xc} have been confirmed in Ref.~\cite{Buehler:2013fha}.
\item Very recently, the full scale dependence of the N$^3$LO expression has
  been constructed in Ref.~\cite{Buehler:2013fha}.
\item The triple-real contribution to the gluon-induced partonic cross 
  section has been considered in Ref.~\cite{Anastasiou:2013srw}. In particular, a method
  has been developed which allows the expansion around the soft limit,
  i.e. for $y = 1 - M_H^2/s \to 0$. Two expansion terms in $y$ are provided.
\item A building block for third-order Higgs boson production is the NNLO
  correction to the Higgs plus jets production which
  has been considered in Ref.~\cite{Boughezal:2013uia} for the gluon-gluon
  channel.
\item An approximate N$^3$LO expression has been constructed in Ref.~\cite{Ball:2013bra} 
  from the resummation of soft-gluon and high-energy singularities.
  (See also Ref.~\cite{Moch:2005ky} for earlier work along similar lines.)
\end{itemize}

\section*{Acknowledgements}
I would like to thank Maik H\"oschele, Jens Hoff, Alexey Pak, Takahiro Ueda
and Nikolai Zerf for fruitful collaborations on the topics discussed in this
contribution.  This work was supported by the DFG through the SFB/TR~9
``Computational Particle Physics''.

\end{document}